\documentclass[twocolumn]{aastex63}
\usepackage{bm}
\usepackage{enumitem}
\usepackage{amsmath}
\usepackage{float}

\newcommand{\curl}[1]{\nabla \times #1}
\newcommand{\divergence}[1]{\nabla \cdot #1}
\newcommand{\norm}[1]{\left\lVert #1 \right\rVert}

\shorttitle{Post-shock Turbulence}
\shortauthors{Foley et al.}

\begin{document}

\title{Development of Turbulence in Post-shock Regions}

\correspondingauthor{Michael Foley}
\email{mfoley@g.harvard.edu}

\author[0000-0002-6747-2745]{Michael M. Foley}
\affiliation{Center for Astrophysics $|$ Harvard \& Smithsonian, 60 Garden Street, Cambridge, MA 02138}

\author[0000-0001-6631-2566]{Philip Mocz}
\affiliation{Center for Computational Astrophysics, Flatiron Institute, 162 5th Avenue, New York, NY 10010, USA}

\author[0000-0001-5817-5944]{Blakesley Burkhart}
\affiliation{Center for Computational Astrophysics, Flatiron Institute, 162 5th Avenue, New York, NY 10010, USA}
\affiliation{Department of Physics and Astronomy, Rutgers University, 136 Frelinghuysen Road, Piscataway, NJ 08854, USA}

\author[0000-0001-6950-1629]{Lars Hernquist}
\affiliation{Center for Astrophysics $|$ Harvard \& Smithsonian, 60 Garden Street, Cambridge, MA 02138}

\author[0000-0003-1312-0477]{Alyssa Goodman}
\affiliation{Center for Astrophysics $|$ Harvard \& Smithsonian, 60 Garden Street, Cambridge, MA 02138}

\begin{abstract}

Supersonic isothermal turbulence is ubiquitous in the interstellar medium. This work presents high-resolution AREPO hydrodynamical simulations of isolated shocks moving through supersonic turbulence to study the development and evolution of turbulence in the pre- and post-shock regions. We find that shocks can amplify turbulent energy in the post-shock region while inducing a preferential orientation for the vorticity. This results in the creation of anisotropic turbulence in the post-shock region. Turbulent energy and dissipation are also strongly enhanced near the shock front. By applying typical scalings from the cold neutral medium to simulations, we find that shocks moving into turbulence on the scale of superbubbles can generate compressive flows on the order of $10^{3} M_{\odot}$/Myr. Our results also show good agreement with related studies on turbulent fluctuations generated by shocks in pure fluid mechanics. 

\end{abstract}


\section{Introduction} \label{sec:intro}

Supersonic turbulence can be found at nearly every scale in astrophysics, ranging from the intergalactic medium to protostellar disks. Within molecular clouds, supersonic turbulence is pivotal for driving gas fragmentation and gravitational collapse into protostars. Turbulence is now known to be a key component for models of many facets of the star formation process, such as the star formation rate \citep{Krumholz2005, Padoan2012, Federrath2012, Padoan2014, Murray2015, Burkhart2018} or the initial mass function \citep{Padoan2002, Ballesteros-Paredes2007, Pudritz2013, Hennebelle2019}. As a result, a large body of literature has emerged that constrains the global statistics of turbulence, such as the power spectrum, structure function, and bispectrum. \citep{Kritsuk2007, Schmidt2008, Federrath2009, Burkhart2009, Price2010}. Understanding the structure of turbulence and shocks moving into turbulence is critical for understanding star formation. The highest resolution simulation of supersonic turbulence to date was achieved with a grid-based code by \citet{Federrath2021}. This work constrained the statistics of isotropic supersonic turbulence in exquisite detail. In the present work, we seek to complement the existing literature on isotropic supersonic turbulence with an investigation of anisotropic turbulence produced by an isolated shock.

Many astrophysical phenomena produce isolated shocks that may interact with upstream interstellar supersonic turbulence, including stellar winds, supernovae, and AGN. Despite their ubiquity, these interactions (versus the structure of turbulence globally) have only recently received increased focus in the astrophysical literature \citep[e.g.][]{Ohlin2019, Bennett2020, Beattie2021, Hu2022Bfield, Hu2022Velocity, Davidovits2022, Rigon2023}. However, shock-turbulence interactions have been well-studied in the fluid dynamics community. Several fluid dynamics studies find that, under an adiabatic equation of state, shocks passing through isotropic subsonic turbulence can strongly enhance the turbulent kinetic energy just downstream of the shock \citep{Larsson2013, Riyu2014, Quadros2016, Tian2017, Braun2019, Boukharfane2018}. This enhancement then decays farther into the post-shock region. Multiple studies also find that vorticity oriented transverse to the direction of shock propagation is enhanced right after a shock \citep{Riyu2014, Livescu2016, Tian2017, Boukharfane2018}. This enhanced transverse vorticity then decays due to viscous dissipation, while streamwise vorticity (oriented along the direction of shock propagation) quickly increases to equilibrate with the transverse vorticity. It is important to note that, in fully developed turbulence, there is no preferential direction for vorticity. Rather, studies such as \citet{Livescu2016} and  \citet{Boukharfane2018} note that post-shock hydrodynamic turbulence is no longer 3D isotropic turbulence; it becomes locally 2D and axisymmetric within the post-shock region. The turbulence returns to 3D as the flow evolves from the shock. 

These findings from the fluid dynamics community are important in several astrophysical contexts, such as understanding the characteristics of dense, star-forming gas, shocks from supernovae, accretion shocks, etc. Yet, these studies have only focused on shocks with relatively low Mach numbers moving through subsonic turbulence. To effectively apply such results to astrophysical phenomena, we must confirm that they hold for stronger shocks and supersonic turbulence. 

Motivated by astrophysical applications, we conduct a parameter study of 3D isolated shock-turbulence simulations. We use a variety of Mach numbers to study the shock-turbulence interaction and the characteristics of post-shock regions in compressible turbulence. While future work should include gravity and magnetic fields, this paper neglects both to focus on the hydrodynamics in dense regions generated by shocks. The only varying parameters in this study are the shock mach number ($\mathcal{M}_{s}$), the turbulence mach number ($\mathcal{M}_{t}$), the turbulence type (decaying or driven), and the simulation resolution. It is also important to recall that the results in the fluid mechanics literature rely on a non-isothermal equation of state. The simulations here are strictly isothermal to facilitate comparisons with simulations of isothermal turbulence in the astrophysical literature. 

We organize the paper as follows: Section~\ref{sec:arepo} discusses the setup of 3D simulations of a shock moving through supersonic turbulence using the moving-mesh code {\sc Arepo}; Section~\ref{sec:interpolation} presents the interpolation and fitting scheme used to identify the shock front and post-shock regions in the simulations; In Section~\ref{sec:shock_results}, we present the results from the shock simulations, including turbulence enhancement, turbulence anisotropy, and converging flows generated by shocks; In Section~\ref{sec:star_formation}, we present brief implications of our results for star formation theory, followed by our conclusions in Section~\ref{sec:conclusions}.

\section{Shock and Turbulence Simulations} \label{sec:arepo}
\subsection{Simulations}

\begin{table}
\begin{center}
\begin{tabular}{ccccc}
\hline
\hline
label & $\mathcal{M}_{s}$ & $M_{t}$ & Turb. Type & Resolution \\
\hline
M5\_T2 & 5        & 2 & driven & 
$512\times128^{2}$ \\
M5\_T5 & 5        & 5 & driven & 
$512\times128^{2}$ \\
M10\_T2 & 10        & 2 & driven & 
$512\times128^{2}$ \\
M10\_T5 & 10        & 5 & driven & 
$512\times128^{2}$ \\
M15\_T2 & 15        & 2 & driven & 
$512\times128^{2}$ \\
M15\_T5 & 15        & 5 & driven & 
$512\times128^{2}$ \\
M15\_T10 & 15        & 10 & driven & 
$512\times128^{2}$ \\
M5\_T2\_R256 & 5        & 2 & driven & $1024\times256^{2}$ \\
M10\_T2\_R256 & 10        & 2 & driven & $1024\times256^{2}$ \\
M15\_T2\_R256 & 15        & 2 & driven & $1024\times256^{2}$ \\
\hline
M5\_T2\_decay & 5        & 2 & decaying & $512\times128^{2}$ \\
M5\_T5\_decay & 5        & 5 & decaying & $512\times128^{2}$ \\
M10\_T2\_decay & 10        & 2 & decaying & $512\times128^{2}$ \\
M10\_T5\_decay & 10        & 5 & decaying & $512\times128^{2}$ \\
M15\_T2\_decay & 15        & 2 & decaying & $512\times128^{2}$ \\
M15\_T5\_decay & 15        & 5 & decaying & $512\times128^{2}$ \\
M15\_T10\_decay & 15        & 10 & decaying & $512\times128^{2}$ \\
M5\_T2\_decay\_R256 & 5        & 2 & decaying & $1024\times256^{2}$ \\
M10\_T2\_decay\_R256 & 10        & 2 & decaying & $1024\times256^{2}$ \\
M15\_T2\_decay\_R256 & 15        & 2 & decaying & $1024\times256^{2}$ \\
\hline
\hline
\end{tabular}
\end{center}
\caption{Simulation parameters. $\mathcal{M}_{s}$ denotes the Mach number of the shock, $\mathcal{M}_{t}$ denotes the Mach number of the pre-shock turbulence, and Turb. Type indicates if the turbulence in the box is allowed to decay naturally or if the turbulence is driven continuously.}
\label{tbl:shock_sims}
\end{table}

\begin{figure*}[!ht]
\centering
\hspace{1.75cm}
\includegraphics[width=0.7\textwidth]{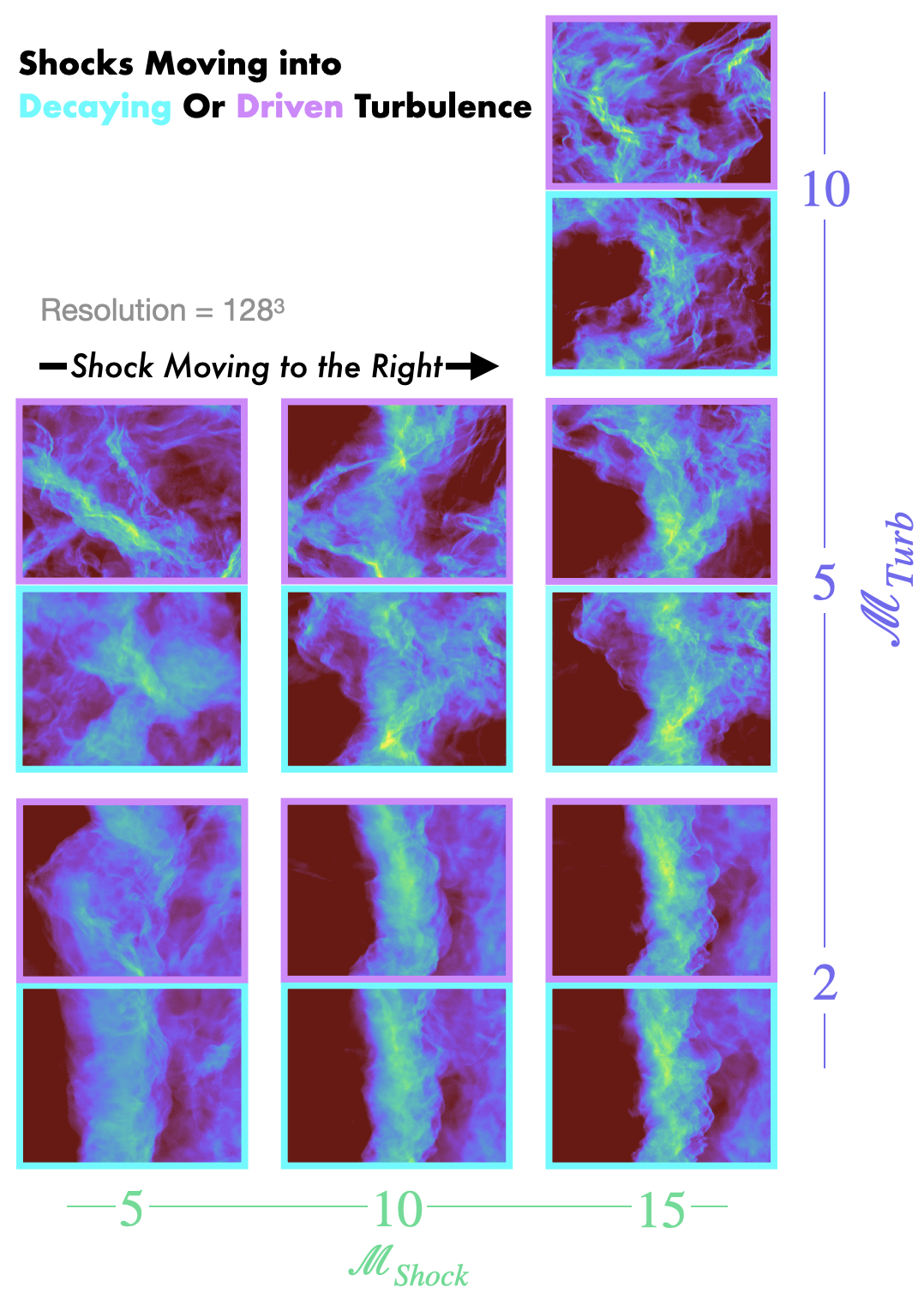}
  \caption{Density projections of shocks moving into either driven turbulence (purple outline) or decaying turbulence (blue outline) at $512\times128\times128$ resolution. Increasing shock Mach number $\mathcal{M}_{s}$ is shown from left to right, and increasing turbulent Mach number $\mathcal{M}_{t}$ is shown from bottom to top. All snapshots are taken when the shock front/remaining density enhancement reaches a location of 0.5 in code units while propagating to the right. Shocks moving into decaying turbulence generally maintain a more coherent structure for longer.} 
  \label{fig:128_Panel}
\end{figure*}

We conduct a suite of 3D simulations of an isolated shock moving through compressible turbulence. We consider both decaying and driven turbulence for this analysis. The parameters of the runs can be seen in Table \ref{tbl:shock_sims}. The simulations are performed using the moving-mesh code {\sc Arepo} \citep{Springel2010}. This quasi-Lagrangian code can resolve high-density contrasts with minimal advection errors, allowing for the resolution to be preferentially allocated to dense post-shock regions while maintaining a high degree of accuracy. 

As mentioned above, our simulations only include hydrodynamics; no gravity, MHD, or other physics are included in this work. Consequently, the simulations are characterized by only three variables: shock Mach number ($\mathcal{M}_{s}$), turbulence Mach number ($\mathcal{M}_{t}$), and type of turbulence (decaying or driven). They are run using an elongated grid setup in {\sc Arepo}. We apply periodic boundary conditions with dimensions of the box set to $(L_{x}, L_{y}, L_{z}) = (4, 1, 1)$ in code units. 

All simulations start with a uniform density of $\rho$ = 1, and we drive turbulence in the elongated box from $6.28 \leq k \leq 12.57$ for at least five turbulence crossing times (measured using the Mach number of the turbulence and a box length of 1). Here $k$ refers to the wavenumber of the turbulence driving with a prefactor of $2\pi$. The kinetic energy reaches equilibrium within roughly one turbulent crossing time. The turbulence driving is purely solenoidal and performed using an Ornstein-Uhlenbeck process \citep{Bauer2012}. After five crossing times, we introduce a shock propagating in the x-direction by assigning a Gaussian velocity pulse to the center of the box. However, since this is a periodic box, we are free to shift the box such that the shock injection occurs at the left edge. With the sound speed set to unity, the resulting x-velocity profile of the box is described by:

\begin{equation}
  v_{x}(x) = \left\{
     \begin{array}{@{}l@{\thinspace}l}
        \mathcal{M}_{s} \times \exp\left(-\frac{(10x-2)^2}{2}\right) &: [0, 0.4] \\
        u_x^{'}(x) &: (0.4, 4.0] \\
        
     \end{array}
   \right.
\end{equation}

where $\mathcal{M}_{s}$ is the Mach number of the shock and $u^{'}_{x}$ denotes the unchanged turbulent x-velocity fluctuations induced by driving. 

The native resolution is $512\times128 \times128$ for the boxes of length  $(L_{x}, L_{y}, L_{z}) = (4, 1, 1)$. It must be noted, though, that {\sc Arepo} is a moving mesh code, so the $512\times128 \times128$ merely represents the starting configuration of the mesh. As the gas evolves, the mesh deforms with the gas, yielding much higher resolution in dense regions and much lower resolution in diffuse regions. In the shock region, we tend to obtain an average resolution of over 1.5 times higher than the fiducial resolution, and a maximum resolution of nearly 4 times higher ($2048\times512\times512$ equivalent). We also run six simulations at a higher resolution of $1024\times256\times256$ to test convergence. Simulations are initially run for one shock crossing time (measured using the initial $\mathcal{M}_{s}$ and the elongated box length). If the shock persists after one crossing time, we extend the simulation for another crossing time. However, we stop the simulation if the shock front encounters the resulting rarefaction wave due to periodic boundary conditions. Consequently, all simulations measure only the effects of a shock moving into pristine supersonic turbulence. $\mathcal{M}_{t}$ is allowed to vary below $\mathcal{M}_{s}$, but simuslations with $\mathcal{M}_{t} >= \mathcal{M}_{s}$ almost immediately destroyed the shock and are not considered in the majority of the analysis (with the exception of M5\_T5).

\begin{figure}[h!]
\includegraphics[width=0.5\textwidth]{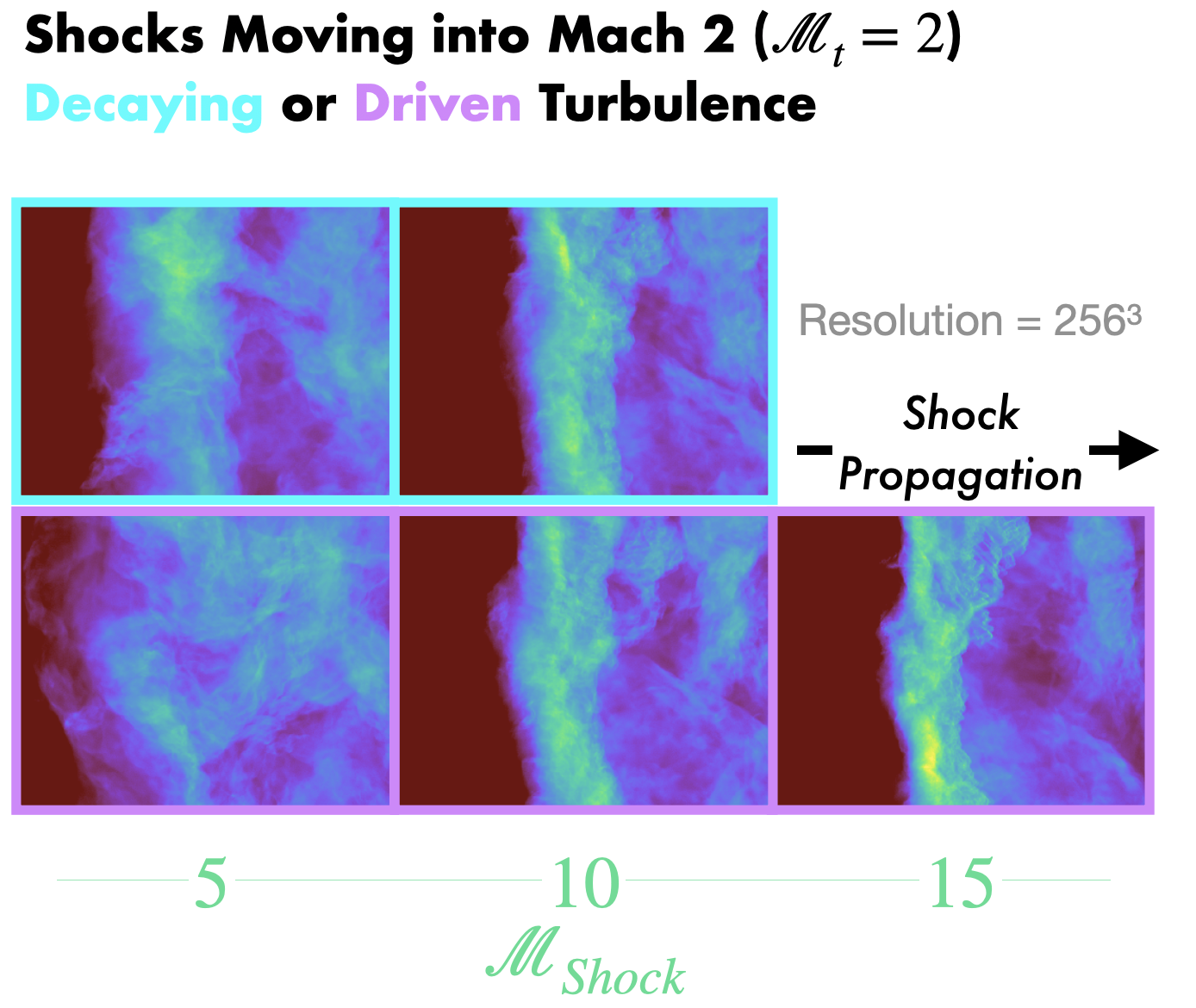}
  \caption{Density projections of shocks moving into both decaying and driven turbulence at $1024\times256\times256$ resolution. Increasing shock Mach number $\mathcal{M}_{Shock}$ is shown from left to right, and all simulations are run with $\mathcal{M}_{t} = 2$. All snapshots are taken when the shock front/remaining density enhancement reaches a location of 0.5 in code units with the shock propagating to the right. Comparing the $1024\times256\times256$ panels to Figure \ref{fig:128_Panel} shows that resolution does not appear to have a major effect on the topology of the shock, at least at early times.} 
  \label{fig:256_Panel}
\end{figure}

In the driven turbulence case, random velocity fluctuations occur throughout the box, including within the shock front. To simulate decaying turbulence, we turn off the turbulence driving as soon as the shock has been injected into the box. The turbulence takes some time to decay, but the shock will not be subject to the addition of random velocity fluctuations to its velocity profile as in the driven turbulence case. 

\begin{figure*}[ht!]
\centering
\includegraphics[width=0.9\textwidth]{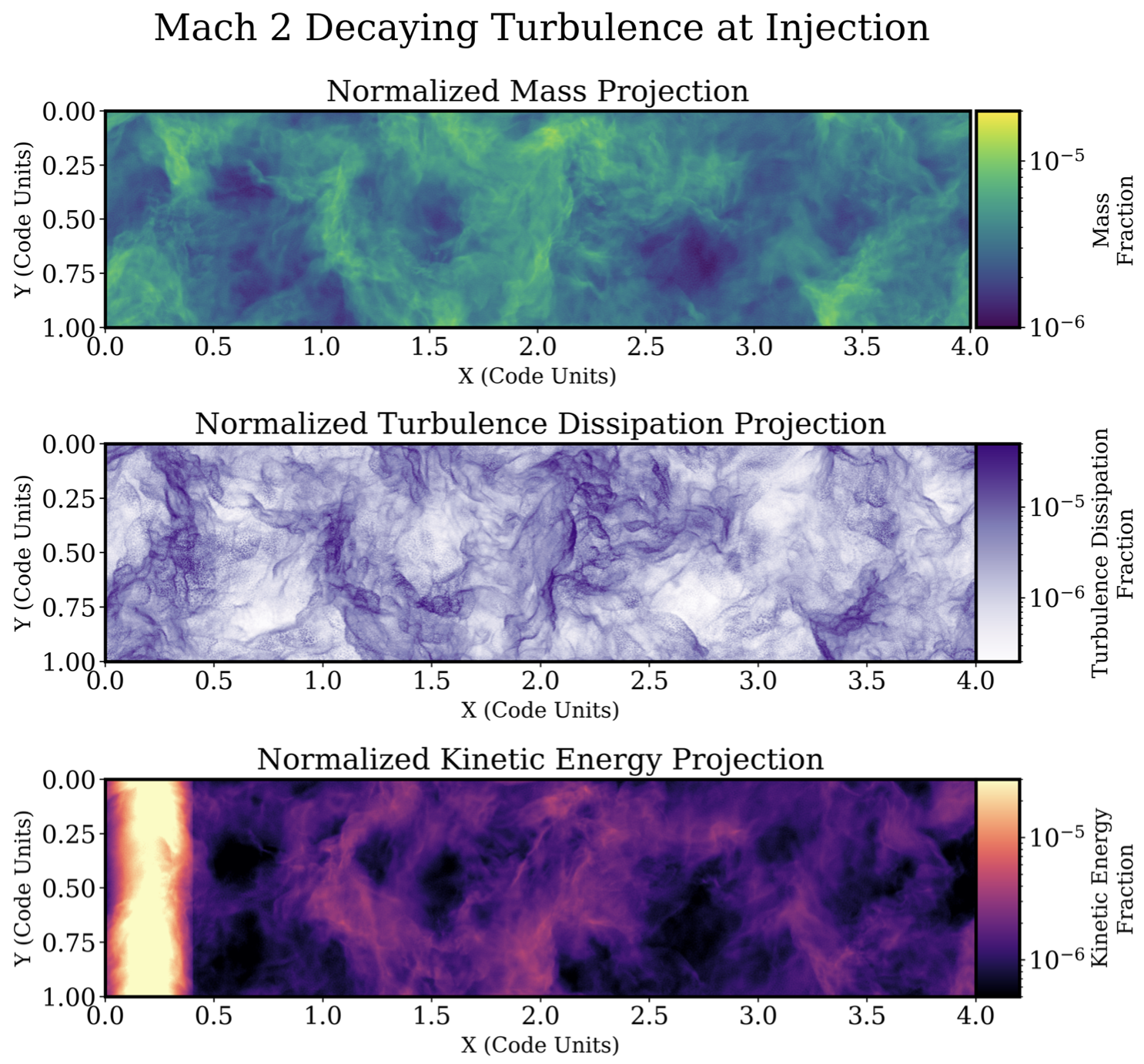}
  \caption{\textit{Top}: Mass projection of the turbulent conditions for the 512x128x128 resolution runs with Mach 2 turbulence after five crossing times. At this snapshot, a velocity pulse is injected to produce a shock.
  \textit{Middle}: Projection of energy dissipated throughout turbulence in the simulation box for the same snapshot. Dissipation is the strongest in the densest regions, though the dissipation does not perfectly track density.
  \textit{Bottom}: Kinetic energy in the box at the moment of shock injection. A Gaussian velocity pulse is injected at the left of the box. Turbulence remains in the rest of the box, and kinetic energy generally matches the mass distribution.} 
  \label{fig:IC_Box}
\end{figure*}

In Figures \ref{fig:128_Panel} and \ref{fig:256_Panel}, we show density projections at fiducial resolution of all of the simulations featured in this work. The projections are computed when the shock front (or remaining density enhancement) has crossed 0.5 code units in the box. Many shocks propagating into the turbulence of $\mathcal{M}=5$ or $\mathcal{M}=10$ dissipate soon afterwards. Indeed, lifetimes of shocks propagating in driven turbulence with a Mach number greater than roughly $\mathcal{M}_{s}/2$ appear quite short. Lifetimes are discussed more quantitatively in Section \ref{subsec:lifetimes}. 

As expected, shocks propagating into decaying turbulence remain more coherent and survive longer. The longer lifetimes are likely due to the lack of random velocity fluctuations added to the shock velocity profile that can disrupt the shock front in the driven case, not a substantial difference in the upstream turbulence. However, it is much easier for the shock to propagate through decaying turbulence versus actively driven turbulence once the pre-shock has had time to decay.

Figure \ref{fig:256_Panel} shows similar projections at higher resolution ($1024\times256\times256$), which are used to test for convergence. Qualitatively, higher resolution does not appear to significantly affect the topology of the shocks. Stronger small shocks are observed near the shock front, but otherwise, the density enhancements appear to be consistent with the lower-resolution runs. 

Before injecting the shock, we first drive turbulence for five crossing times to ensure that the turbulence reaches a fully developed state. For example, projections of the mass, dissipation, and kinetic energy of $\mathcal{M}_{t} = 2$ turbulence in Figure \ref{fig:IC_Box} show the initial conditions for the $M_{t} = 2$ runs as the shock is injected. Minor periodicity in the x-direction can be observed due to the driving process in an elongated box, but this should not influence the results in a significant way due to the fully developed nature of the turbulence. In the kinetic energy plot, the pulse at the left shows the initial condition of the shock injected into the box.

Strong turbulent dissipation is observed in regions of high density in our simulations, supporting the model from \citet{Boldyrev2002_KBModel} that proposes, in supersonic compressible turbulence, energy decays in the inertial range via vortices and dissipates via shocks. However, the correspondence is not precisely one-to-one, and this discrepancy becomes more apparent after the injection of the shock into the turbulent box.

\begin{figure*}[ht!]
\centering
\includegraphics[width=\textwidth]{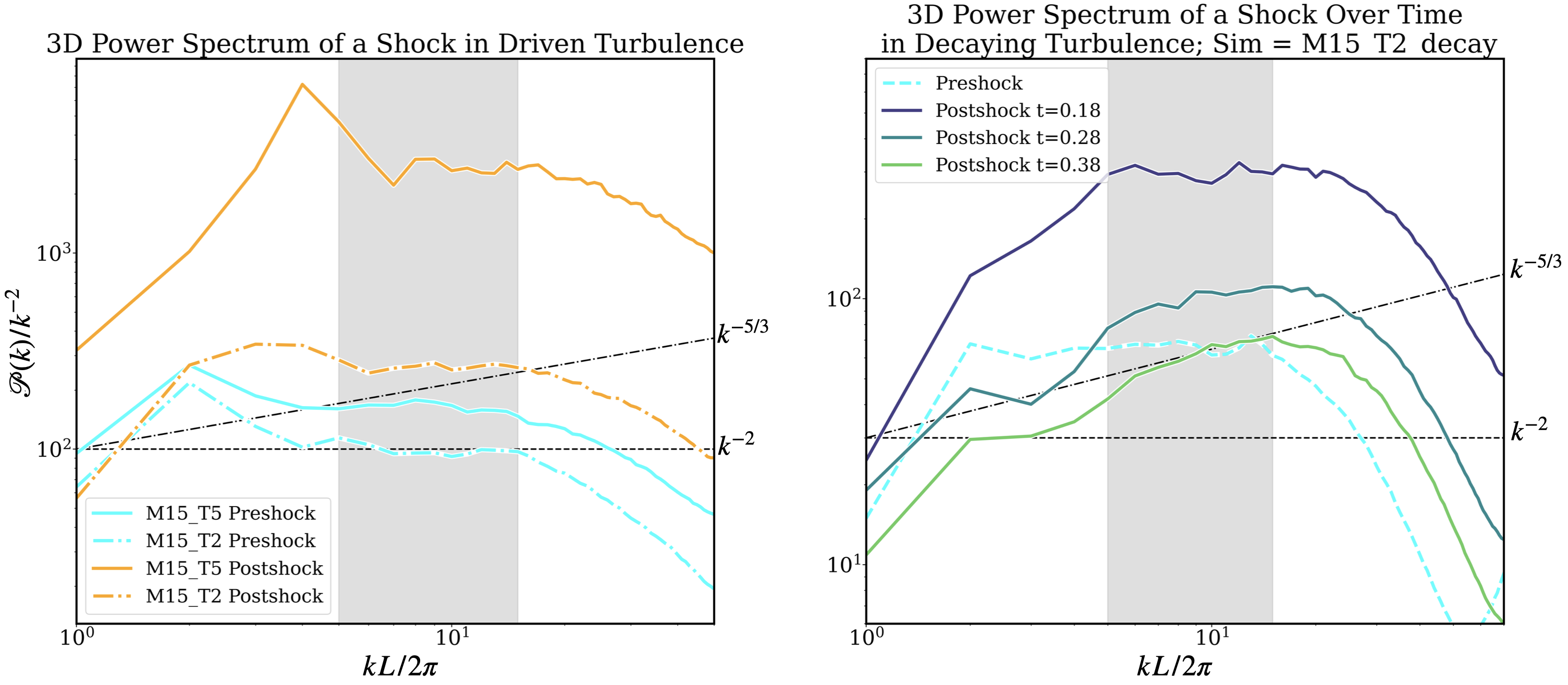}
  \caption{\textit{Left}: 3D kinetic energy power spectrum of the pre-shock and post-shock regions for a Mach 15 shock passing through either Mach 2 or Mach 5 driven turbulence. $L$ is the characteristic length of the box. The power spectra are divided by $k^{-2}$ to aid visualization. The gray region roughly indicates the inertial range of the spectra where the slopes should be evaluated. All spectra appear to follow a Burgers scaling of $k^{-2}$. 
  \textit{Right}: 3D kinetic energy power spectrum of three different time snapshots for a Mach 15 shock passing through either Mach 2 decaying driven turbulence. At early times, the shock appears to follow $k^{-2}$ scaling but decays to $k^{-5/3}$ more quickly than the surrounding turbulence.} 
  \label{fig:3d_ps}
\end{figure*}

To ensure our turbulence is fully developed, we compute the kinetic energy power spectrum in the pre-shock turbulence for all of our boxes. We follow \citep{Kritsuk2007} and compute the 3D velocity power spectrum as:

\begin{equation}
    \mathcal{E}(\mathbf{k}) = \frac{1}{2}\bigl|\mathbf{\tilde{u}(k)}\bigr|^2.
\end{equation}

\begin{equation}
    \mathcal{E}(k)\;\equiv\;\int_{V}\mathcal{E}(\mathbf{k})\;\delta\bigl(\lvert\mathbf{k}\rvert - k\bigr)\;d\mathbf{k}\,.
\end{equation}

where $\mathbf{k}$ is the wavenumber, $\mathbf{u}$ is the velocity, and $\mathbf{\tilde{u}}$ is the Fourier transform of the velocity. As we are selecting regions that remain periodic in the $y$ and $z$ axes but not in the $x$ axis, we applying a Hanning window to our data to smooth out non-periodic artifacts at the boundaries of our data. We then can measure the spectrum in the inertial range of the turbulence to understand how energy moves from large scales to small scales. Kolmogorov turbulence is known to have a scaling of $k^{-5/3}$, while Burgers, or shock-dominated, turbulence is known to have a scaling of $k^{-2}$. In Figure \ref{fig:3d_ps}, we evaluate which model better captures the turbulent dynamics present in different regimes of our simulation. The spectra reveal that inertial range ends by $kL/2\pi \approx 20-30$, which corresponds to resolutions of 10-20 moving mesh cells in each dimension within the densest post-shock regions.

We find that our pre-shock regions maintain Burgers scalings throughout their evolutions, only flatting the slopes of their inertial ranges at late times in the decaying runs. The shock regions do appear to exhibit additional power at lower scales than the pre-shock regions, indicating some larger-scale driving of turbulence distinct from the general driving of turbulence throughout the box. The post-shock regions also maintain Burgers scalings through the inertial range for most of their evolutions, but simulations with strong shocks moving into decaying turbulence show interesting deviations. We calculate that, as a shock moves through weaker decaying turbulence, the shock tends to adopt a Kolmogorov scaling before the ambient turbulence (Figure \ref{fig:3d_ps}, right panel). In other words, the shock appears to aid the transition to incompressible turbulence as energy decays through the system. The mechanisms that might underpin such a transition are discussed more in Section \ref{subsec:HOG}.

As our simulations are scale-free, we can also choose any characteristic length scale, sound speed, and density to convert to physical units. To scale velocities and lengths, the reader can multiply the appropriate quantities by the sound speed or length scale. For masses, one must use the assumed length scale and an assumed density to convert the code units into physical ones. From these scalings, all other quantities of interest can be derived. When showing physical results in this paper, we adopt typical values for a molecular cloud: a length scale $L$ of 5.2 pc (which yields a box size of 5.2 pc x 5.2 pc x 20.8 pc), a sound speed $c_{s} = 0.2~{\rm km}~{\rm s}^{-1}$, and a number density $n_{H} =$ 100 $\rm {cm^{-3}}$. A number density of $n_{H} =$ 100 $\rm {cm^{-3}}$ with a mean particle mass $\langle m \rangle = 1.4m_{\textrm{H}}$ yields a total $M \approx 2 \times 10^{3} ~M_{\odot}$ for the box. Such scalings yield a mass resolution of $m = 3 \times 10^{-5} ~M_{\odot}$ per cell. This fiducial scaling corresponds to the mass of more diffuse portion of a molecular cloud, which allows us to investigate the dynamics of shock-induced compression. Otherwise, we report results in code units so that the reader can scale the simulations appropriately to their use case. 

While the hydrodynamics are truly scale free, it is important to note that other physics affecting the gas are certainly not. Thermal physics, gravity, and magnetic fields will all affect the hydrodynamics differently at different scales. Our simulations neglect all of these contributions to first constrain the effects of purely isothermal hydrodynamics. Furthermore, while our choices of sound speed, length scale, and density correspond to independent scales in the hydrodynamics, they are strongly coupled in the real interstellar medium. The choice of density and sound speed defines a thermodynamic state, and the choice of length scale introduces a dynamical time that must be compared to the cooling time of the gas. 

The domain of realism for our simulation scalings is, therefore, first restricted to cases where the isothermal approximation is valid: gas where the cooling time is short compared to the dynamical time. Neglecting self-gravity also requires the Jean's length to remain large compared to the chosen length scale of the simulation. Finally, a small beta is needed to allow for magnetic fields to be neglected. No single phase of the ISM perfectly satisfies all of these criteria. However, large-scale shocks in dense, cold gas or small-scale shocks in warm, diffuse gas represent the areas of the ISM best captured by our simulations. Values for $L$, $c_{s}$, and $n_{\textrm{H}}$ characteristic of large-scale regions in molecular clouds, the Cold Neutral Medium, or small-scale regions within the Warm Neutral Medium are therefore the most realistic scaling choices for our simulations. Other choices may lead to contradictions within various observational regimes.

\section{Interpolation and Shock Fitting}
\label{sec:interpolation}

As mentioned before, {\sc Arepo's} moving mesh means our simulation snapshots do not possess a fixed grid. However, much of our analysis relies on a fixed grid, meaning that we must interpolate onto a Cartesian mesh. Several algorithms for Cartesian interpolation exist, but we choose to use an SPH interpolation algorithm to preserve many of the benefits of the moving mesh. Our code implementation draws from the tools used in the \verb|meshoid| package developed by Michael Grudi\'c \citep{Grudic2021}. The algorithm mimics the original SPH formulation of \citet{Springel2005}, and we interpolate two conserved quantities --- mass and momentum --- using the procedure described below.

First, we establish a Cartesian mesh of a fixed resolution. We choose the resolution of the Cartesian mesh to be double the native resolution of the Arepo simulation. For instance, a simulation run at $128\times128\times512$ resolution in Arepo is interpolated onto a fixed Cartesian grid of $256\times256\times1024$. This is done to ensure that the Cartesian mesh captures as many of the high-resolution regions as possible. Once the grid is defined, we use the smoothing length of each moving mesh cell to select the Cartesian grid cells into which we will deposit the conserved quantity. After specifying the deposition domain, we calculate the weight to apply to each Cartesian cell using the spline kernel for SPH schemes \citep{Monaghan1985, HK1989, Springel2005}:

\begin{equation}
W(r,h) = \frac{8}{\pi h^{3}}\left\{
     \begin{array}{@{}l@{\thinspace}l}
        1 - 6\biggl(\dfrac{r}{h}\biggl)^{2} + 6\biggl(\dfrac{r}{h}\biggl)^{3} &: 0 \leq \dfrac{r}{h} \leq \dfrac{1}{2} \\[10pt]
        2\biggl(1 - \dfrac{r}{h}\biggl)^{3} &: \dfrac{1}{2} < \dfrac{r}{h} \leq 1 \\[10pt]
        0 &: \dfrac{r}{h} > 1
     \end{array}
   \right.
\end{equation}
where $h$ is the smoothing length

\begin{equation}
h = (\textrm{Moving Mesh Cell Volume})^{1/3}\\
\end{equation}
and $r$ is the distance from the center of the moving mesh cell to the center of the Cartesian cell.

We then iterate over every cell in the moving mesh, calculating the contribution of each cell to the surrounding Cartesian cells. Finally, we normalize the kernel weights to ensure that the deposition conserves the initial quantity. Ultimately, the density ($\rho$) of the conserved quantity ($q$) in each Cartesian cell is given by 

\begin{equation}
\rho_{i} = \sum_{j=1}^{N} q_{j}W(|r_{ij}|, h_{j})
\end{equation}

where $N$ is the number of moving mesh cells and $i$ denotes each Cartesian cell after weight normalization for each moving mesh cell.

After interpolation, we identify the x-coordinates marking the boundary of the post-shock region, the peak of the shock front, and the first density enhancement due to the shock (as the shock front is not a perfect discontinuity due to the turbulence into which it is propagating). We only use the x-momentum ($p_{x}$) data to identify the bounds and peak of the shocked region. Our fitting algorithm proceeds with the following steps:
\begin{enumerate}
    \item Use a Gaussian filter to smooth $p_{x}$.
    \item Compute the 1D profile of $p_{x}$ in the direction of shock propagation.
    \item Find the main peak of the profiles using the \verb|find_peaks| function from \verb|scipy| \citep{Scipy}.
    \item Identify left and right boundaries in $p_{x}$ of the main peak, calculated using $1.5\times $FWHM of the main peak. The extra factor of 1.5 ensures we capture most of the dense post-shock region and the start of the density enhancement in front of the shock.
    \item Plot location of shock boundaries and shock peak across snapshots to view shock evolution over simulation time.
    \item Fit 4th degree polynomial to shock evolution, where earlier times (where the shock is strongest) are weighted most heavily. This fitting aims to reduce noise in the selection procedure and yield a smooth shock evolution. We find that a 2nd degree polynomial effectively captures the bulk of the shock evolution, but a 4th degree polynomial allows for better fitting of early and late times in the shock evolution.    
\end{enumerate}

Each shock simulation will then have a single 4th-degree polynomial that can be used to compute the shock peak and boundary locations at every timestep. In the next section, we use the polynomial fits to calculate all results. These fits compute the boundaries indicated by black dashed and solid lines in subsequent plots.

\begin{figure*}[ht!]
\centering
\includegraphics[width=0.9\textwidth]{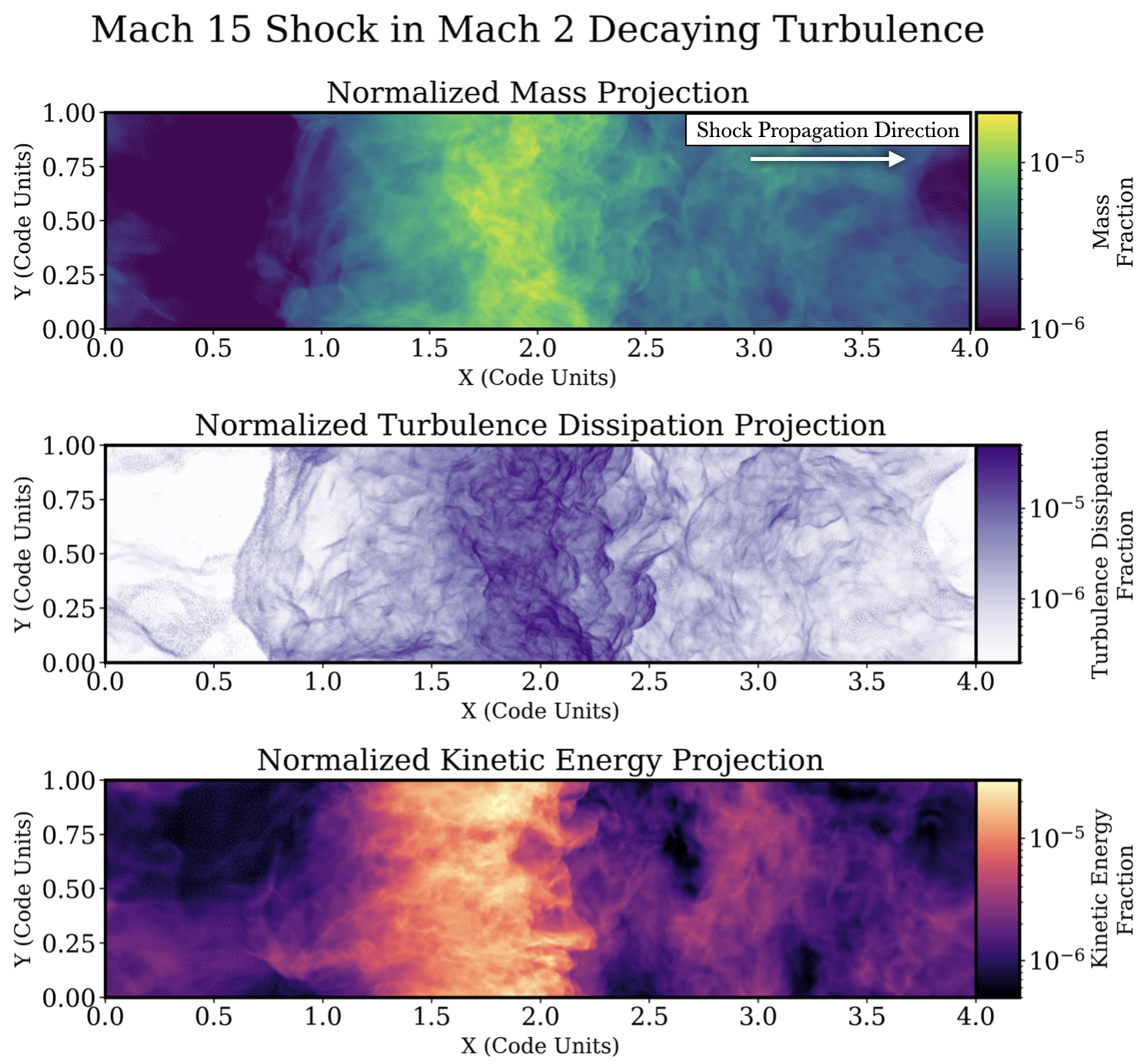}
  \caption{\textit{Top}: Mass projection of a shock with $\mathcal{M}_{s} = 15$ and $M_{t} = 2$ after one shock crossing time (where a crossing time is defined as $t_{\rm cross} = L_{\rm box}/\mathcal{M}_{s, {\rm initial}}$) for a 512x128x128 run. The low-density region on the left of the plot is the rarefaction wave, and the shock is traveling to the right.
  \textit{Middle}: Projection of energy dissipated by turbulence for the same snapshot. Dissipation is observed to be the strongest in the dense regions, though it does not perfectly track density. 
  \textit{Bottom}: Projection of kinetic energy for the same snapshot. As the shock propagates, it develops protrusions of high kinetic energy surrounded by regions of kinetic energy closer to the ambient turbulence.}
  \label{fig:Density_TurbDiss_Shock}
\end{figure*}

\section{Results} \label{sec:shock_results}
We consider five sets of diagnostics from the simulations:

\begin{enumerate}[topsep=0pt]
    \setlength{\itemsep}{1pt}
    \setlength{\parskip}{0pt}
    \setlength{\parsep}{0pt}
    \item Turbulent Energy and Dissipation 
    \item Vorticity and Divergence of Velocity
    \item Converging and Diverging Momentum Flows
    \item Histogram of Oriented Gradients (HOG) for Velocity
    \item Shock Lifetimes

\end{enumerate}

Section \ref{subsec:turb_energy} discusses how shocks affect the dissipation and turbulent energy production in the simulations. Section \ref{subsec:vorticity} looks at the vorticity and compressive motions throughout the simulation box, particularly focusing on anisotropies just behind the shock front. Section \ref{subsec:HOG} presents the calculation of the Histogram of Oriented Gradients, a statistical technique we use to measure orientations of gas motion, comparing the results from pre-shock and post-shock regions. Section \ref{subsec:converge} considers the converging and diverging flows of momentum behind the shock and discusses their implications for star formation. Section \ref{subsec:lifetimes} presents the lifetimes of shocks moving into turbulent gas with our adopted scalings. For clarity and brevity (and unless otherwise noted), the plots in this paper showing a single shock-turbulence interaction will all pertain to a single run: M15\_T2\_decay. In all projection plots shown, the shock is moving to the right.

\subsection{Turbulent Energy and Dissipation} \label{subsec:turb_energy}

For isothermal simulations, turbulent energy can be defined as the total kinetic energy minus the kinetic energy of the fluid's bulk motion. In the case of a purely turbulent box, the bulk motion of the fluid is zero, and the total turbulent energy of the box is simply the total kinetic energy. As viscosity and diffusion remove energy from the box, turbulent dissipation in a given time step can be measured by subtracting the initial kinetic energy of the box from the kinetic energy at the end of that timestep. 

In Figure \ref{fig:Density_TurbDiss_Shock}, we again plot projections of density and energy dissipation with a shock injected into the box. A few prominent features are immediately visible. First, the turbulence ahead of the shock has been allowed to decay as the shock propagates. Consequently, density enhancements and turbulent energy dissipation are less pronounced than in the pristine turbulence of Figure \ref{fig:IC_Box}. The shock strongly enhances energy dissipation due to the creation of smaller turbulent and dissipative structures, which we analyze in greater detail in the following sections. The edge of the rarefaction shock experiences strong dissipation, even with low density, due to free-streaming gas encountering a significant jump in density. Finally, the shock develops protrusions of high kinetic energy in its front, suggesting that the turbulence creates ``holes'' around these protrusions of lower velocity gas. Their kinetic energy appears comparable to the pristine turbulence. 

It is also interesting to note that turbulence dissipation rises before the kinetic energy, with strong dissipation beginning around $x = 2.25$ and the bulk of kinetic energy arriving around $x = 2.0$. We can conceptualize this dissipation by investigating the turbulent energy distribution around the shock. If the dissipation increases but the turbulent energy does not, then the front of the shock will quickly become laminar. However, if both dissipation and turbulent energy increase, the shock can maintain an energetic equilibrium and serve as a persistent, dissipative structure as it moves through the turbulence. Throughout the entire sample of shock Mach numbers and turbulent Mach numbers, we observe consistent increases in both dissipation and turbulent energy behind the shock front. However, the shock fronts never become fully laminar, suggesting that a nonuniform shock front may enhance turbulent energy as much, or more, than it dissipates.

\begin{figure}[ht!]
  \gridline{\includegraphics[width=0.47\textwidth]{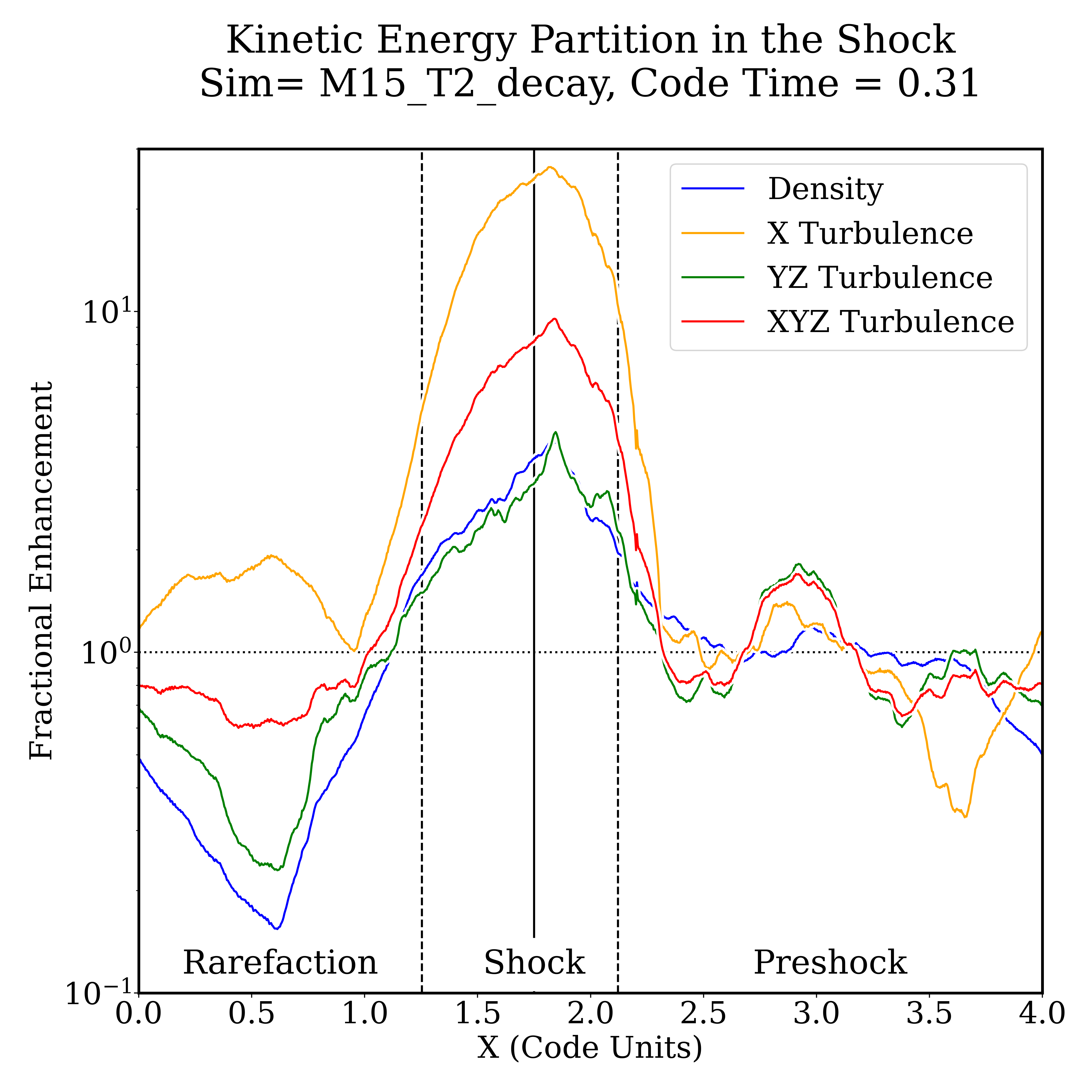}}
  \caption{Log density and energy profiles for a single snapshot of the M15\_2\_decay run. The density profile is shown in blue, the total turbulent energy is shown in red, turbulent energy in the $y$ and $z$ directions (fluctuations parallel to the shock front) is shown in green, and turbulent energy in $x$ (fluctuations in the direction of shock propagation) is shown in yellow. Each profile is expressed as the ratio of the value at a given box index to the median value across the whole pre-shock.}
  \label{fig:Turb_Energy}
\end{figure}

Amplification of the turbulent energy is seen throughout the shock region, rising at the first onset of density enhancement that corresponds with the onset of strong dissipation. The amplification appears to be driven almost entirely by additional turbulent energy generated in the $x$ direction (direction of shock propagation). We can observe this by looking at 1D profiles of the energy in Figure \ref{fig:Turb_Energy}. 

In Figure \ref{fig:Turb_Energy}, we show the breakdown of turbulent energy across the shock profile. We subtract the mean velocity profile (weighted by mass) from each of the three velocity dimensions to yield only the fluctuating, turbulent components of the velocity field: $v^{'}_{x}$, $v^{'}_{y}$, and $v^{'}_{z}$. We then square each component and multiply by mass to measure the total turbulent kinetic energy in each dimension. We consider three combinations of the dimensions: $m(v^{'}_{x})^{2}$ alone (turbulent motions in the direction of the shock propagation), $m((v^{'}_{y})^{2} + (v^{'}_{z})^{2}))$ (turbulent motions parallel to the shock front), and total turbulent energy $m((v^{'}_{x})^{2} + (v^{'}_{y})^{2} + (v^{'}_{z})^{2})$. These combinations and density are then measured relative to their pre-shock values and plotted as fractional enhancements.

To compare with observations, we recall our fiducial scalings yield a box size of $5.2 ~\textrm{pc} \times 5.2 ~\textrm{pc} \times 20.8 ~\textrm{pc}$. Consequently, the shock region in \ref{fig:Turb_Energy} occupies a physical length of roughly 5 pc. For this snapshot, the total mass within the shock is approximately $M_{shock} \approx 1 \times 10^{3} M_{\odot}$. On our interpolated grid, the shock is resolved by 250 cells along its length in this snapshot, yielding a spatial resolution of 0.02 pc. Such resolution allows us to investigate the kinematics from cloud to clump scales.

\begin{figure*}[ht!]
\hspace{1.5cm}  \includegraphics[width=15cm]{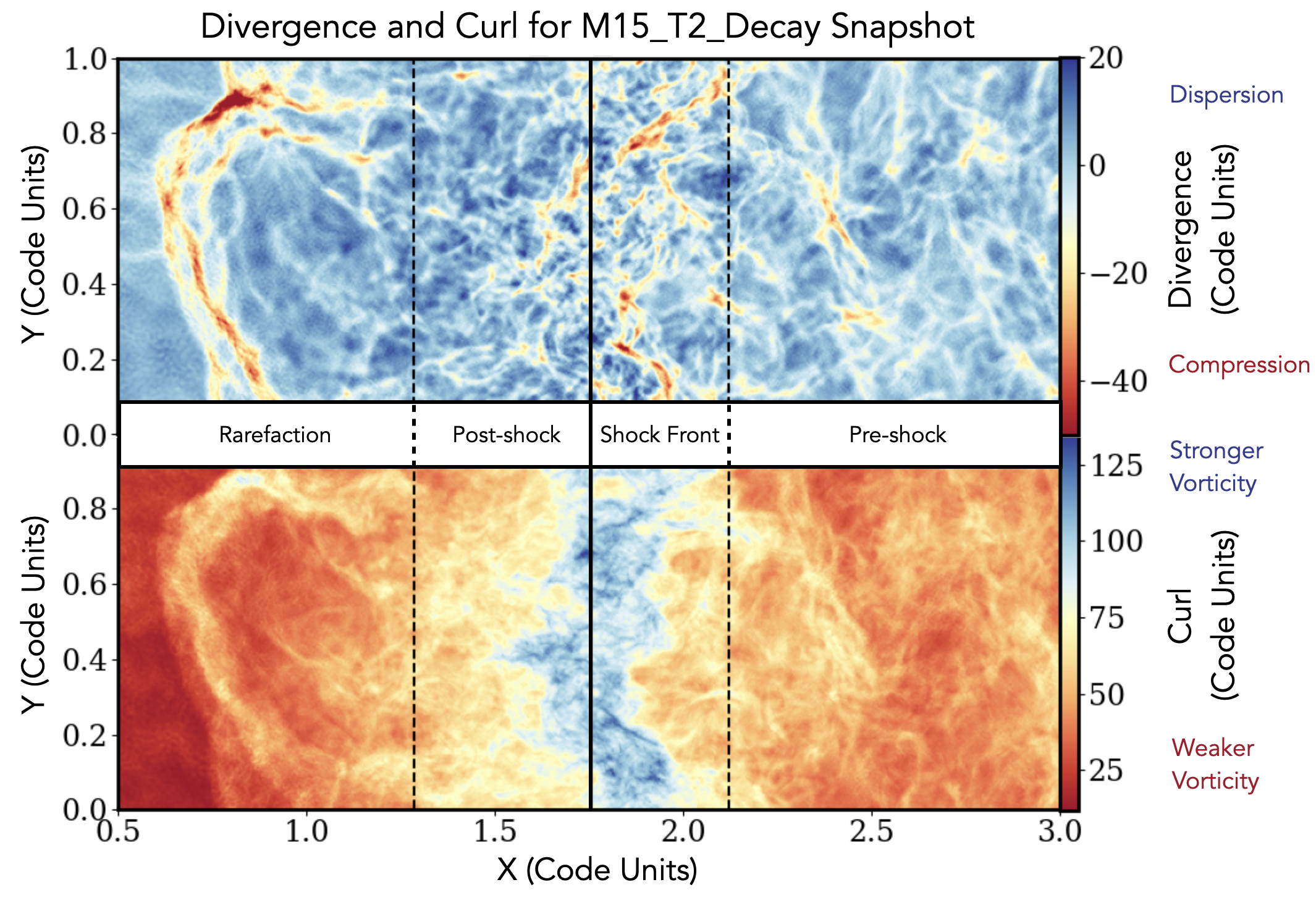}
  \caption{\textit{Top}: Projection of the divergence of the 3D velocity field for a snapshot of the M15\_T2\_decay run. The solid black line indicates the shock density peak, the left black dashed line indicates the scale height of the post-shock region, and the right black dashed line indicates the transition point from pre-shock to shock. Red regions are compressing, and blue regions are dispersing.
  \textit{Bottom}: Projection of curl of the velocity field. As in the top plot, the black lines correspond to the shock region boundaries. Blue regions possess strong vorticity, while red regions are characterized by flow with little rotation.} 
  \label{fig:CurlDiv}
\end{figure*}

The total turbulent energy is strongly enhanced at the shock peak relative to the pre-shock, primarily driven by large increases in the $v^{'}_{x}$ motions. The YZ turbulence energy appears to follow density very closely in the shock region, suggesting that there is not much enhancement of turbulent velocities in this region. Instead, the turbulence appears to maintain whatever $v^{'}_{y}$ and $v^{'}_{z}$ motions where present at the start of the shock front. In the rarefaction region, $v^{'}_{y}$ and $v^{'}_{z}$ motions are suppressed as material tends to stream freely in the x-direction, and overall turbulence is slightly suppressed.

The $v^{'}_{x}$ motions are strongly enhanced as soon as the front of the shock reaches a density enhancement above pre-shock densities. They rapidly increase until the shock peak, at which time they begin to decrease throughout the post-shock region. Once density drops below pre-shock values in the rarefaction region, though, $v^{'}_{x}$ motions are again enhanced due to the propagation of the rarefaction wave and the free-streaming of gas within that region.

Similar enhancements of turbulent fluctuations in the direction of shock propagation are observed in the literature for subsonic upstream turbulence (i.e. see \citet{Quadros2016}, \citet{Tian2019}, \cite{Livescu2016}). The rarefaction shock experiences a similar enhancement, driven by free-streaming gas running into a quickly increasing density profile. Most of the excess energy is dissipated by the end of the density peak. To understand the mechanism for this dissipation, we can examine the vorticity (curl of the velocity field) and divergence of the velocity field throughout the shock and post-shock region.  

\begin{figure*}[ht!]
 \includegraphics[width=18cm]{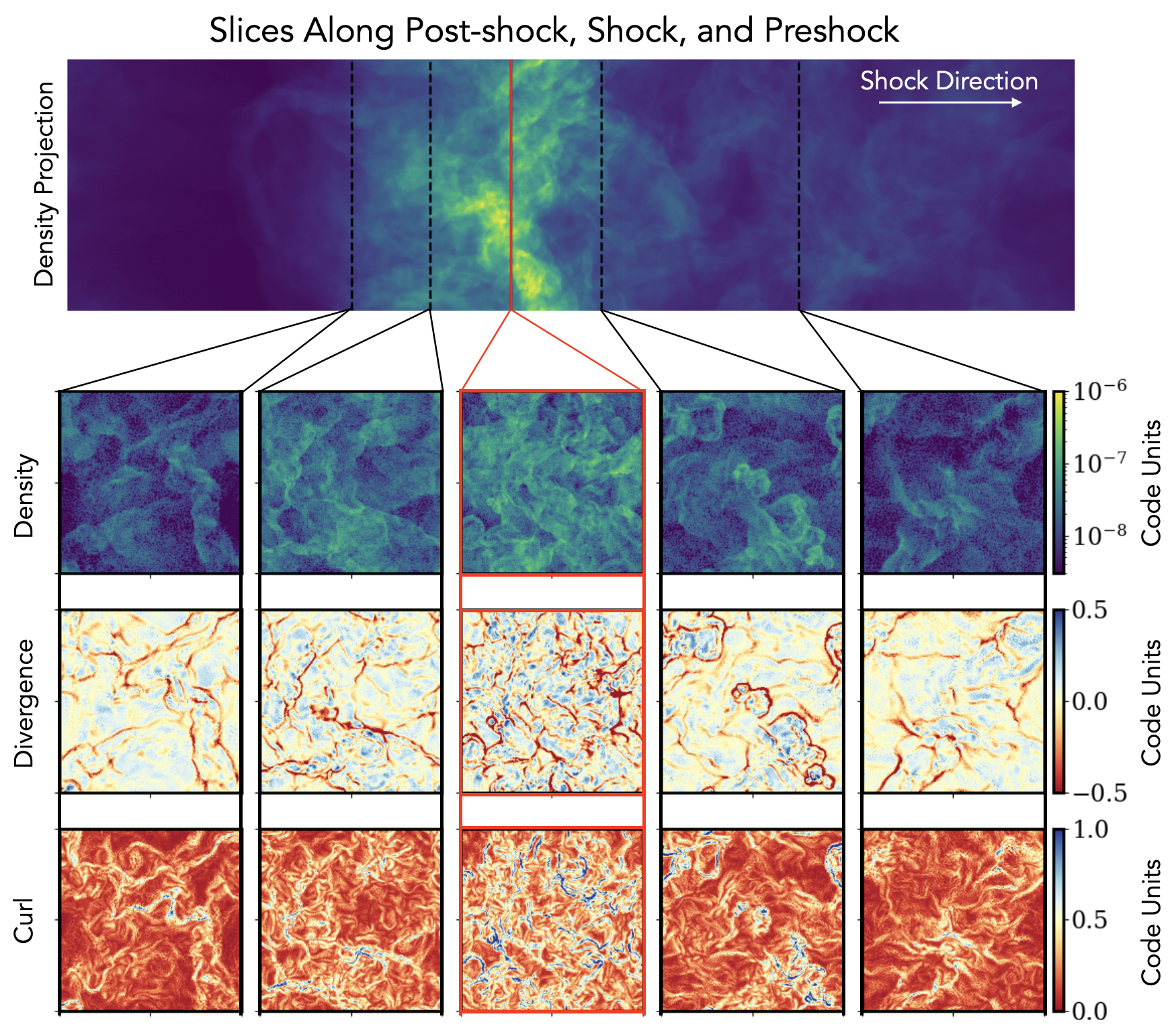}
  \caption{\textit{Top}: Projection of the density for a snapshot of the M15\_T2\_decay run. Different lines indicate different slices taken through the simulation domain where the curl and divergence are calculated. Starting a the right, the first black line indicates pre-shock gas. The second black line indicates the approximate region of the start of the shock. The red line indicates the shock density peak. The third black line indicates the scale height of the post-shock region. Finally, the fourth black line shows the end of the post-shock region.
  \textit{Top Slices}: Density at each slice in the simulation domain. Density is given in code units, which can be scaled to physical units using the scalings introduced earlier in the paper.  
  \textit{Middle Slices}: Divergence of the velocity field at each slice in the simulation domain. Negative values for the divergence indicate compressive motions, while positive values indicate diffusive ones.
  \textit{Bottom Slices}: Curl of the velocity field at each slice in the simulation domain. Larger values indicate stronger vorticity, while smaller values indicate more laminar flow.} 
  \label{fig:DensityDivCurl}
\end{figure*}

\subsection{Vorticity and Divergence of Velocity} \label{subsec:vorticity}

The driving of turbulence can be broadly decomposed into two categories: solenoidal modes ($\divergence{v} = 0$) and compressive modes ($\curl{v} = 0$). Consequently, we can learn about the nature of the turbulence in the post-shock region by studying both the curl and the divergence of the velocity field. Figure \ref{fig:CurlDiv} and \ref{fig:DensityDivCurl} show the curl and divergence calculated throughout a single snapshot of the M15\_T2\_decay run. Since the divergence and curl only rely on velocity, certain structures (like the rarefaction wave) are more prominent.

In Figure \ref{fig:CurlDiv}, the shock front is characterized by motions with a large curl magnitude and a strongly negative divergence. The large magnitudes of both the vorticity and the divergence suggest that turbulence is driven with both solenoidal and compressive motions at the shock front and immediately behind it, contributing to the strong turbulent energy enhancement observed in Figure \ref{fig:Turb_Energy}. While the post-shock region exhibits a higher vorticity magnitude and a higher compression than the pre-shock region, both compression and vorticity approach equilibrium shortly after the post-shock scale height.

One feature of note is the relatively clumpy distribution of both the vorticity and divergence, especially near the shock front. Such ``clumps" with a strong negative divergence are found in a high-density region, which could serve as the precursors of star-forming clumps. Additionally, due to the relatively high turbulent energy in the post-shock region, turbulent ``pressure" may help to confine the compressive structures and form analogs to ``droplets" \citep{Chen2019}. This discussion is continued in Section \ref{sec:star_formation}.

\begin{figure}[ht!]
\gridline{\includegraphics[width=0.45\textwidth]{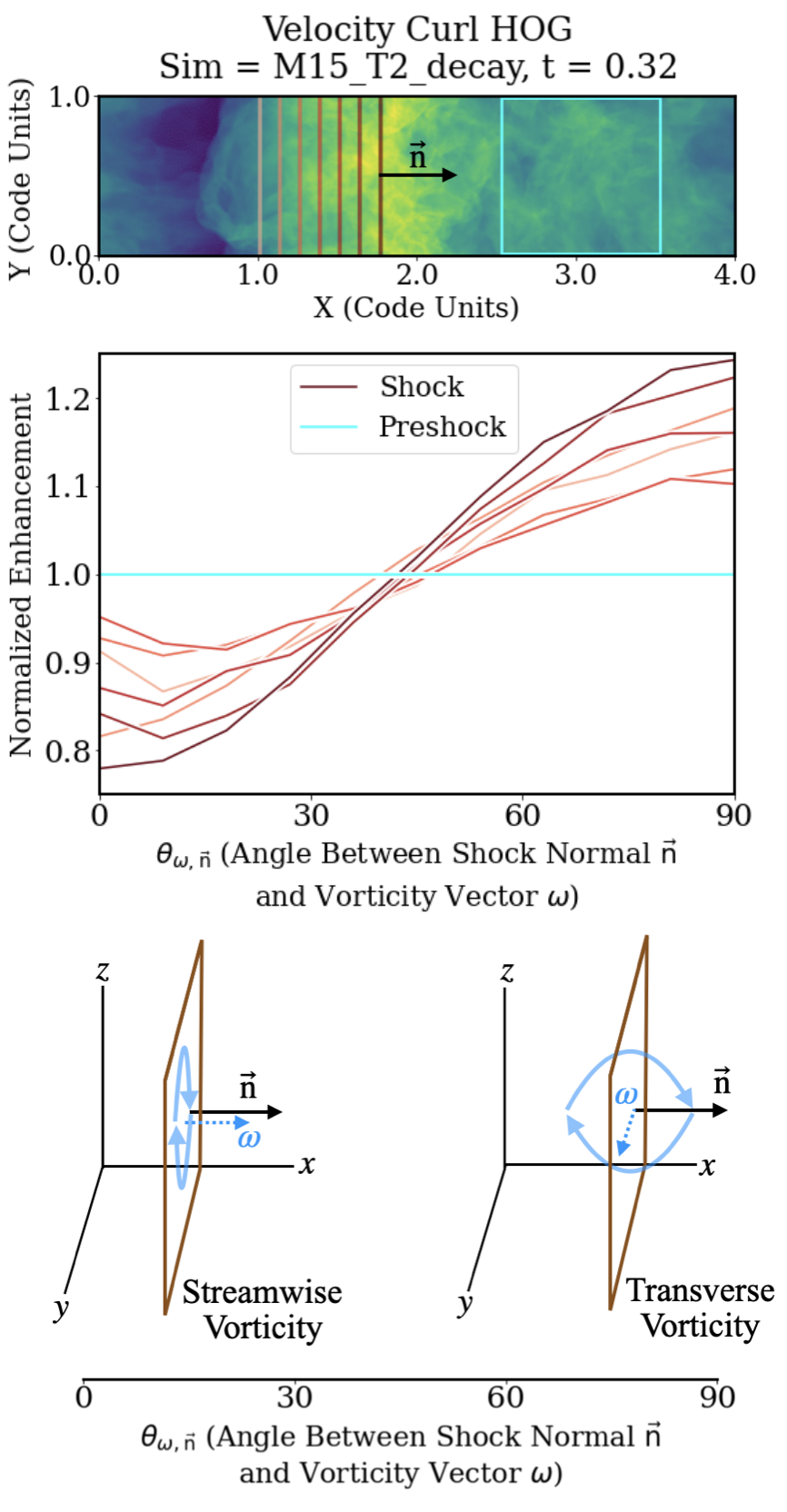}}
  \caption{\textit{Top}: Projection of turbulent dissipation for the same snapshot as shown in Figure \ref{fig:Density_TurbDiss_Shock}. Lines overlaid show the slices over which the Histogram of Oriented Gradients was computed in the post-shock region. The black vector indicates the shock normal vector.
  \textit{Middle}: Histogram of Oriented Gradients for the vorticity field. For every slice in the top plot, the angle between the vorticity vector and the shock normal is calculated. Counts of each angle are binned, tallied, and reported as a histogram for each region. Counts are normalized to the mean pre-shock value, which is divided out to highlight deviations.
  \textit{Bottom}: Schematic showing the different vorticity orientations that are enhanced or depressed in the HOG. Low angles between $\omega$ and $\vec{\rm{n}}$ correspond to streamwise vorticity and high angles correspond to transverse vorticity.}
  \label{fig:HOG_Vorticity}
\end{figure}

\begin{figure*}[ht!]
  \gridline{\includegraphics[width=18cm]{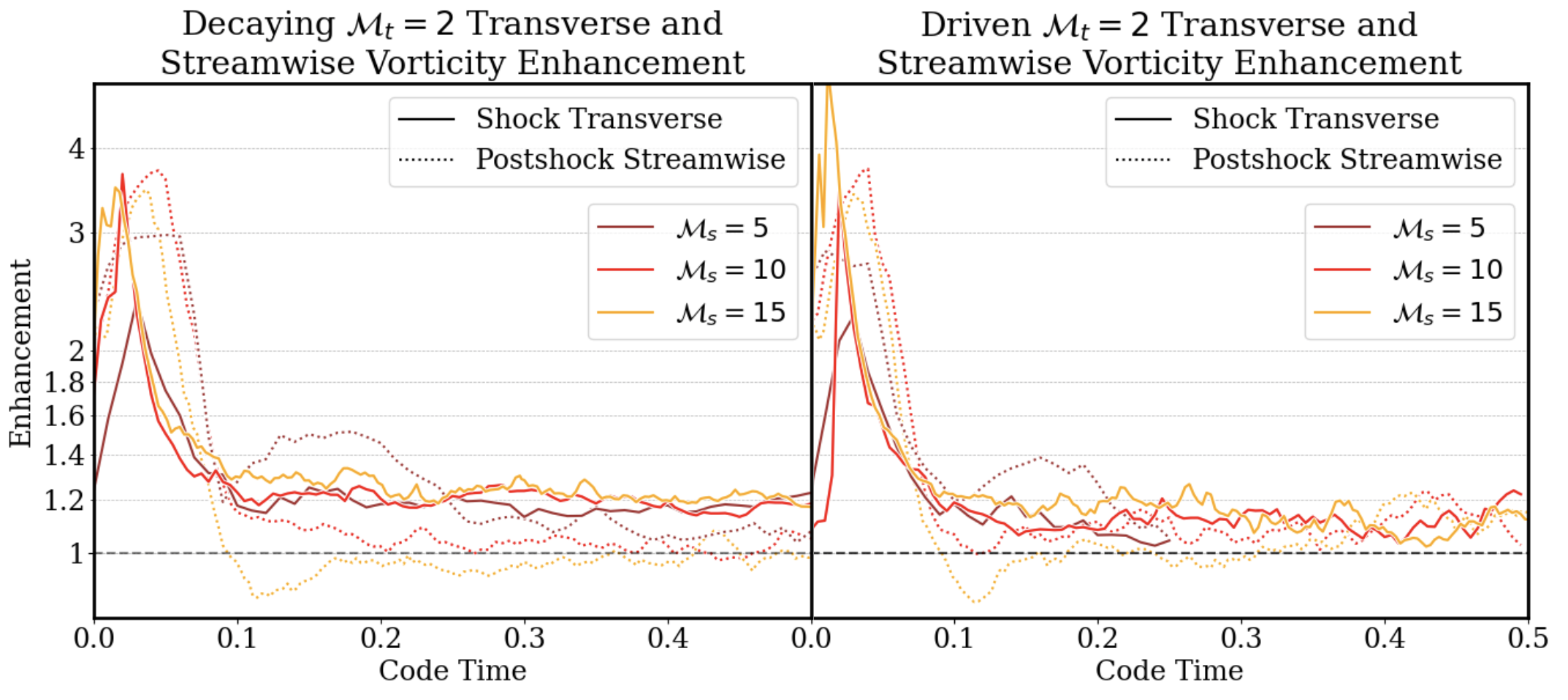}}
  \caption{\textit{Left}: Streamwise and transverse vorticity enhancement in the shock region for runs with decaying $\mathcal{M}_{t} = 2$ turbulence. The y-axis is shown as a log scale with reference values provided. Solid lines indicate transverse vorticity measured at the shock front, and dotted lines indicate streamwise vorticity measured at the end of the post-shock region. Different colors indicate different runs where the shock Mach number $\mathcal{M}_{s}$ is varied.
  \textit{Right}: Same as left panel but for simulation runs with driven $\mathcal{M}_{t} = 2$ turbulence.} 
  \label{fig:VorticityEnhancement}
\end{figure*}

Strong compression can also be observed at the rarefaction shock, though the vorticity is much lower than in the shock due to high-velocity, laminar flow. Here, gas can free-stream through the low-density environment of the rarefaction region, eventually running into the higher-density environment of the post-shock region. At this point, a large ``filament" of compressive motions is formed that persists throughout the entire simulation time for all tested parameter combinations. This region is still quite diffuse, though -- nearly an order of magnitude less dense than the shock front. Consequently, despite the strong compressive motions, such a filament appears not possess the necessary mass to initiate gravitational collapse. Instead, it appears to serve primarily as a dissipative structure. Figure \ref{fig:DensityDivCurl} shows the evolution of the divergence and vorticity as gas starts in the turbulent pre-shock, moves through the shock front, mixes with post-shock material, and finally returns to a turbulent state approximating the pre-shock. 

As gas first encounters the corrugated shock front, compression occurs in small, confined regions. These regions also show an increase in vorticity, indicating the gas just upstream of the density peak is already experiencing solenoidal and compressive turbulence motions. Both driving mechanisms correspond with the rise in $v^{'}_{x}$ just before the shock front in Figure \ref{fig:Turb_Energy}. 

As the gas moves into the shock density peak, compressive and solenoidal motions only increase. However, they are no longer found in enhancements with the same level of coherence. Instead, they are distributed much more randomly throughout the shock front and appear more fragmented. Visually, this stage shows the cascade of turbulent energy initially generated at the start of the shock front. The energy cascades down to dissipative vortices in the post-shock region that equilibrate the turbulent energy with pre-shock levels. Moving deeper into the post-shock, longer and more compressive structures develop. Finally, the gas reaches the end of the post-shock and returns a divergence and a curl distribution similar to the pre-shock.

Once gravity is incorporated, the long, compressive regions may form the seeds for bound structures. As the simulations are scale free, larger scale motions in the ISM are also represented by our results. In other words, the formation of filaments or clouds may be seeded by shock-driven compressive motions like these in the ISM. Recent simulation results have shown similar structures emerge in turbulence with and without magnetic fields \citep{Federrath2021, Zhao2024}. In Section \ref{subsec:converge}, we consider estimates of converging and diverging mass flows obtained from applying physical scalings to the divergence maps to evaluate whether our compressive structures might collapse in the presence of additional physics and how such compressive structures relate to observed clouds.

\subsection{Histogram of Oriented Gradients and Power Spectra of Velocity}
\label{subsec:HOG}

Vorticity is a vector, so we can also examine its orientation relative to other structures in the domain. In particular, we consider the orientation of vorticity in the post-shock region relative to the shock front. To do so, we use a technique known as the Histogram of Oriented Gradients (HOG). The HOG originated in the field of computer vision and has recently been used to great effect in studies of the interstellar medium \citep{Soler2019, Beuther2020}. For our application, we define a vector $\vec{n}$ normal to the shock front and compute the relative orientation between the vorticity and the shock front as follows:

\begin{equation*}
\theta_{\omega, \vec{\rm{n}}} = \cos^{-1}(\frac{|\omega \cdot \vec{n}|}{\norm {\omega}\norm{\vec{\rm{n}}}})  
\end{equation*}
where $\vec{\omega}$ is the vorticity and $\vec{n} = \vec{e_{1}}$, a unit vector in the $x$-direction, since the shock purely propagates in that dimension.

Next, we bin the results and plot a histogram to evaluate trends in orientation between the two quantities. Figure \ref{fig:HOG_Vorticity} shows the resulting Histogram of Oriented Gradients (HOG) evaluated at seven different regions throughout the post-shock and averaged throughout the pre-shock region. In isotropic turbulence, the HOG should return a flat profile since all orientations of vorticity are equally likely. We find that our pre-shock profile is very nearly isotropic, though it features slight enhancement in streamwise vorticity vs transverse vorticity. This slight anisotropy is likely due to the driving mechanism employed in our extended box. To correct this in our analysis, we divide out the pre-shock HOG from each post-shock HOG line.

Each of the seven lines in post-shock regions is evaluated at a different location in the post-shock, illustrated by the projection at the top of Figure \ref{fig:HOG_Vorticity}. In general, lighter colors are found deeper into the post-shock region, and darker colors are found closer to the shock front. Just behind the shock front, the HOG indicates that the vorticity prefers to orient itself transverse to the shock normal. In other words, vortices will develop almost like ``wheels" rolling in the direction of shock propagation. Deeper into the post-shock region, streamwise vorticity (with a vorticity vector parallel to the direction of shock propagation) rises to transition the turbulence back to isotropy.

\begin{figure*}[ht!]
  \includegraphics[width=0.99\textwidth]{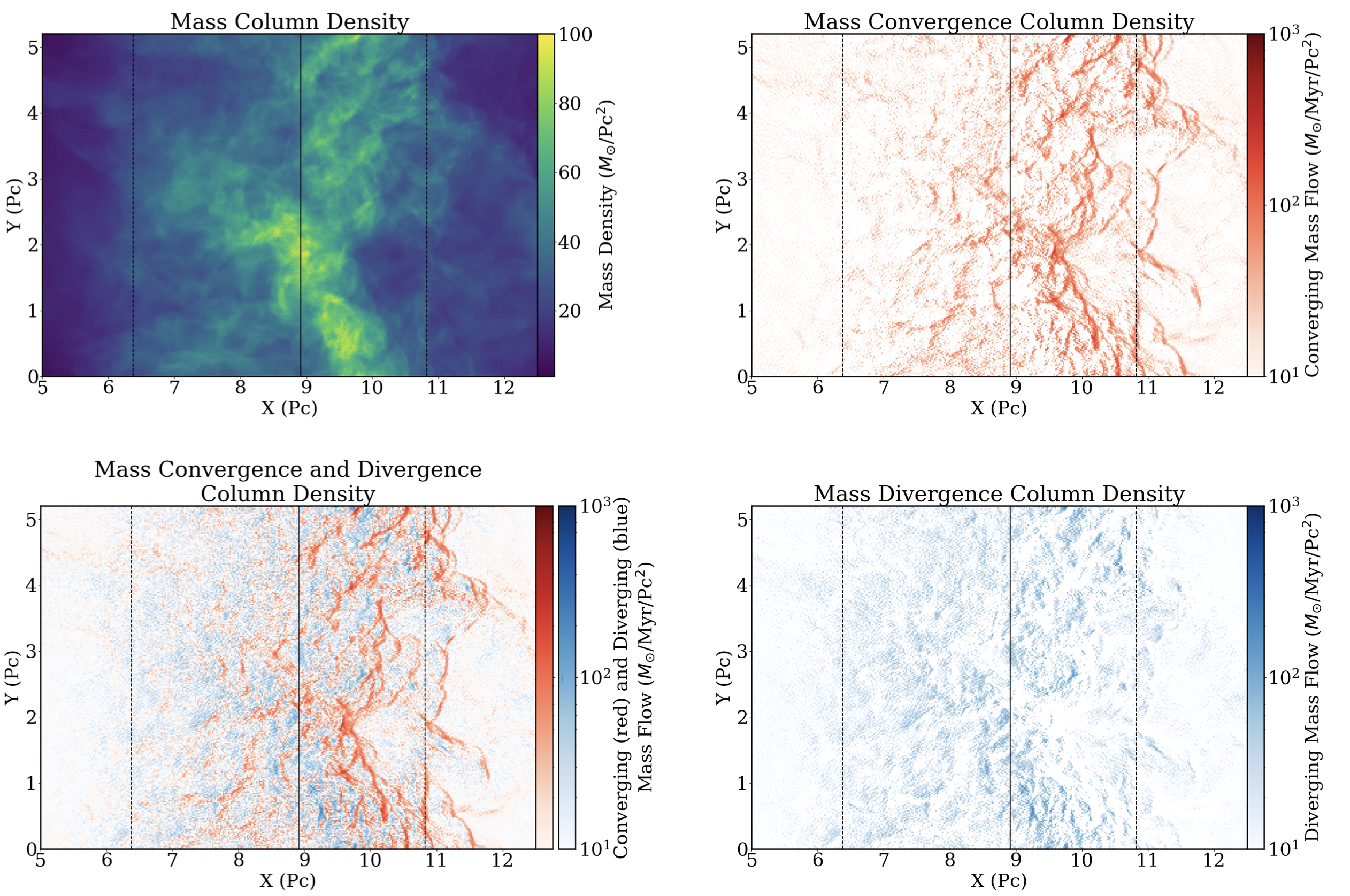}
  \caption{\textit{Top Left}: Mass column density for the shock region in a snapshot of the M15\_T2\_decay run. The dotted lines indicate the bounds of the shock front and post-shock regions, and the solid black line indicates the shock peak.
  \textit{Top Right}: Projection of negative divergence (convergence) of the velocity field in units of column density per Myr ($M_{\odot}/\rm{Myr}/\rm{Pc}^{2}$).
  \textit{Bottom Right}: Projection of positive divergence of the velocity field in the same units as above.
  \textit{Bottom Left}: A composite of the top right and bottom right plots, with projections of positive divergence and negative divergence overlaid. Converging regions are indicated in red, and diverging regions are indicated in blue. The range of the mass flow is the same for both converging and diverging motions.}
  \label{fig:ConvergeDiverge}
\end{figure*}

This picture of enhanced transverse vorticity behind the shock front also confirms what has been found in the literature for a shock with low $\mathcal{M}_{s}$ moving into subsonic turbulence ($\mathcal{M}_{t} < 1$) \citep{Boukharfane2018}. Variations in the degree of anisotropy are primarily due to the evolution of the shock front; at earlier times, the enhancement of transverse vorticity at the shock front can reach a factor of a few compared with the pre-shock region, and streamwise vorticity is also strongly enhanced in the post-shock region. Figure \ref{fig:VorticityEnhancement} shows enhancements in both transverse and streamwise vorticity throughout the shock. The amplitude of transverse vorticity enhancement is evaluated at the shock front, and the amplitude of streamwise vorticity enhancement is evaluated at the end of the post-shock region. In decaying turbulence, the transverse vorticity enhancement remains relatively constant throughout time, while streamwise vorticity tends to decay to pre-shock values over time. One hypothesis for the constant enhancement of transverse vorticity at the shock front is that the transfer of laminar energy to turbulent energy by the shock proceeds at a rate comparable to the energy dissipation rate. In other words, the roughly constant enhancement in transverse vorticity measures a constant rate of energy conversion by the shock. As the post-shock region grows, though, gas has more time to equilibrate to isotropic conditions. Consequently, the measure of post-shock streamwise vorticity drops to pre-shock levels as time progresses. 

In the simulations with driven turbulence, however, both transverse and streamwise vorticity tend to decay slowly over time to pre-shock values. This decay might imply that turbulence is driven by the shock at a slower rate than turbulence is driven in the ambient gas, leading the shock to equilibrate more quickly with the surrounding isotropic turbulence.

\subsection{Converging Mass Flows} \label{subsec:converge}

Using the physical scalings discussed in the introduction (typical values for a molecular cloud), we can rescale the simulation box to physical units of roughly 20 parsecs x 5 parsecs x 5 parsecs. Doing so means that the post-shock regions considered in this paper are on roughly 5-10 parsec size scales, a factor of a few times larger than the size of a typical "clump" ($\sim 1 $pc). Consequently, the hydrodynamics of these dense regions are likely very relevant to the formation of stars. 

By applying physical scalings to the simulation boxes, the previous results on the divergence and curl of the velocity field can be coupled with the gas mass to derive plots showing where mass is converging and where it is diverging. Figure \ref{fig:ConvergeDiverge} shows both the converging and diverging mass flows throughout a shock simulation box. Once gas moves into the shock region, it quickly experiences strong converging motions before the density peak of the shock. Shortly after, diverging motions also rise and serve to return the gas to a state of isotropy. However, calculating the total net compression (compressive motions - dispersive motions) in the region between the peak of shock density and the start of the shock (between the solid black line and the right dotted line in Figure \ref{fig:ConvergeDiverge}) for this snapshot yields $\sim 55 ~M_{\odot}/\textrm{Myr}$ Behind the peak of shock density, the shock experiences net dispersive motions with isolated regions of strong convergence. Even without gravity, the movement of a shock through turbulent gas can generate compressive motions on the size and mass scales necessary to potentially create star-forming clumps within molecular clouds.

As the simulations are scale-free, we are free to adopt any sound speed and density (within the aforementioned domains of validity) to obtain physical scalings. Let us choose $c_{s} = 1 ~\rm{km} ~\rm{s}^{-1}$, $n_{H} = 20 ~\rm{cm}^{-3}$, and $L = 30 ~\rm{pc}$, typical values for the Cold Neutral Medium. The simulation box then has dimensions of $30 ~\rm{pc} \times 30 ~\rm{pc} \times 120 ~\rm{pc}$. Our total simulation box mass also increases to 7.5 $\times 10^{4}$ M$_{\odot}$ One way we can apply the scalings to a realistic scenario is by considering a superbubble shock that is moving into turbulent gas. The shell is quite old, so it is firmly in the momentum-driven regime and does not rely on any other source of propulsion. Scaling up our converging mass flow calculation yields a net compression of $\sim 1.8 \times 10^{3}$ M$_{\odot}$/Myr, approaching the mass of a small molecular cloud. These scalings are representative of a portion of the shell of the Local Bubble, which has been shown to include many of the star-forming molecular clouds in the solar neighborhood \citep{Zucker2022_LB}. Our results suggest that hydrodynamics alone can help explain significant portions of the compression necessary to form molecular clouds on the shells of superbubbles.

To evaluate whether or not these compressive regions represent structures that could collapse in the presence of gravity, we compare the Jean's length to the typical sizes of our compressive regions. To account for the nonthermal support provided by the turbulence, we modify the sound speed to represent the total velocity dispersion as $\sigma_{tot} = \sqrt{c_{s}^{2} + \sigma_{NT}^2}$ \citep[e.g.][]{Pillsworth2025}:

\begin{equation}
    \lambda_{J} = \sigma_{tot}\frac{\pi}{(G\rho)^{1/2}}
    \label{eq:jeans_length}
\end{equation}

We consider the strongest areas of mass flow (converging mass flows measured in the top $1\%$ of the simulation volume) for our analysis. Such regions demonstrate clumpy structure, indicating coherent compressive motions. Using our molecular cloud scalings, the average number density in these regions is $n \approx 600 ~\rm{cm}^{-3}$. We also measure an average nonthermal velocity dispersion at $\mathcal{M} = 1.75$, yielding $\sigma_{NT} \approx 0.35 $ km/s. Our adopted $c_{s} = 0.2$ km/s then yields a Jean's length of $\lambda_{J} \approx 2.5 ~\rm{pc}$. As our box is 5.2 pc across, this length comparable to the size of some of the most compressed regions, though in most cases it is slightly larger. Such a result reflects the plausability that hydrodynamics alone could create compressive motions on the scales necessary for collapse, though additional physics (particularly thermal physics and magnetic fields) are necessary to fully evaluate the possibility for collapse behind shock fronts.

\subsection{Shock Lifetimes}
\label{subsec:lifetimes}

By measuring when density and momentum fluctuations produced be the shock reach the amplitude of fluctuations naturally produced by turbulence in the pre-shock, we can determine the time at which a shock has ``dissipated" into the surrounding turbulence. Table \ref{tbl:shock_life} shows results from the determination of shock lifetimes from the full parameter study. Many of the shocks persisted past the end of the simulation box or the end of the allowed simulation time, so our results can only place lower bounds on their lifetimes. 

To convert the shock lifetimes into physical units, we adopted a sound speed of $c_{s} = 0.2$ km s$^{-1}$ and box size of 5.2 pc $\times$ 5.2 pc $\times$ 20.8 pc. The resulting conversion from code time to physical time was $\approx 25$ Myr (code time)$^{-1}$. Thus, to convert the lifetimes to any other scalings, divide the reported lifetimes by 25 Myr and multiply by the appropriate conversion factor for the selected sound speed and length scale. 

For decaying turbulence, we find that the shortest lifetime is $\approx 1$ Myr for M15\_T10\_decay. The lower bounds can extend up to $\approx 16$ Myr for decaying turbulence, though the runs with $\mathcal{M}_{s} = 15$ demonstrate the most coherent shock front at the end of the simulation time. In general, large $\mathcal{M}_{s}$ and smaller $\mathcal{M}_{t}$ corresponds to longer lifetimes. However, it is interesting to note how quickly the shock is destroyed in the M15\_T10\_decay case. Despite having a shock that initially moves faster than the turbulence, the shock is destroyed much quicker than in the case of M5\_T5\_decay.

For driven turbulence, we find that the shortest lifetime is also $\approx 1$ Myr for M15\_T10. This is comparable to M15\_T10\_decay as the turbulence has not had much time to decay yet, nor has significant energy been injected through turbulence yet, so the dynamics of the shock are comparable. However, there is a significant change in the lifetimes of shocks in simulations with $\mathcal{M}_{t} = 5$. Now, the simulations only yield shock lifetimes of 1-3 Myr, as opposed to lower bounds of 16 Myr in the case of decaying turbulence. Increased runtime and simulation domain beyond the lower bounds are expected to reveal much shorter lifetimes than the decaying case as well.

\begin{table}
\begin{center}
\begin{tabular}{ccccc}
\hline
\hline
label & Lifetime (Myr) & Type &  \\
\hline
M5\_T2 & 6.3 & Lower Bound\\
M10\_T2 & 12.6 & Lower Bound \\
M15\_T2 & 15.2 & Lower Bound \\
M5\_T5 & 1.8 & Lifetime \\
M10\_T5 & 1.4 & Lifetime\\
M15\_T5 & 3.2 & Lifetime \\
M15\_T10 & 1.0 & Lifetime \\

\hline
M5\_T2\_decay & 16.0 & Lower Bound \\
M10\_T2\_decay & 16.0 & Lower Bound \\
M15\_T2\_decay & 16.0 & Lower Bound \\
M5\_T5\_decay & 15.3 & Lifetime \\
M10\_T5\_decay & 16.0 & Lower Bound \\
M15\_T5\_decay & 16.0 & Lower Bound \\
M15\_T10\_decay & 1.3 & Lifetime \\

\hline
\hline
\end{tabular}
\end{center}
\label{tbl:shock_life}
\caption{Shock Lifetimes. The lifetimes are given in Myr by assuming our typical scalings ($L = 5.2 \rm{pc}$ and $c_{s} = 0.2$ km s$^{-1}$). ``Lower Bound" means the simulation ended before the shock was destroyed, either from computation timeout or the shock reaching the end of the box. ``Lifetime" means the shock was destroyed before the end of the simulation, so the reported time is a measurement of how long the shock survived. Most shocks moving into decaying turbulence survived past the end of the simulation. Shocks moving into driven turbulence of $\mathcal{M}_{t} = 2$ survived past the end of the simulation, while shocks moving into stronger driven turbulence were dissipated rather quickly.}
\end{table}

\section{Star Formation Implications} 
\label{sec:star_formation}

Both solenoidal and compressive motions can have a large effect on star formation. Solenoidal motions can provide support against collapse, while compressive motions can induce it. In the simulations, we observe that the post-shock region augments both solenoidal and compressive motions as part of the general turbulent energy enhancement. Such compressive motions can form the seeds for star formation, but solenoidal motions and generally increased turbulent pressure could result in the formation of ``droplets".

\citet{Chen2019} first identified droplets as coherent, pressure-confined structures in Ophiucus and Taurus. Droplets have a typical radius of 0.04 pc and a typical mass of 0.4 M$_{\odot}$. While not virially bound, they are confined by ambient pressure and similar to shock-induced features observed in MHD simulations. \citet{Offner2022} further characterized a 3-stage  evolution of cores produced by turbulence. Using our adopted molecular cloud scalings, the clumpy distribution of divergence and vorticity in the post-shock regions of our simulations possesses the same physical scales as the droplets from \citet{Chen2019} and the turbulent cores from \cite{Offner2022}. Thus, we propose that turbulent motions produced by the shock may serve as the precursors to such structures by compressing the gas and keeping it pressure-confined. Given our mass resolution of $m = 3 \times 10^{-5} ~M_{\odot}$, these droplet-like structures are well-resolved in our simulation, suggesting that hydrodynamics alone can do much of the work to create such structures. Further work is needed at higher resolution, particularly with the addition of magnetic fields and gravity, to investigate this connection.

The critical density for star formation within a molecular cloud can be computed with the following equation \citep{Krumholz2005}: 

\begin{equation}
\rho_{\rm crit}/\rho_{0} = \frac{\pi^{2}}{15}\alpha_{\rm vir} \mathcal{M}_{s}^{2}
\label{eq:postshock_density}
\end{equation}

where $\rho_{0}$ is the mean density of the cloud, $\alpha_{vir}$ is the virial parameter of the cloud, and $\mathcal{M}_{s}$ is the Mach number. The Mach number is defined as the ratio of gas velocity to local gas sound speed; $\mathcal{M}_{s} > 1$ is supersonic. For $\alpha_{vir} \sim 1$, $\rho_{\rm crit}/\rho_{0} \approx \rm \mathcal{M}_{s}^{2}$. For an isothermal shock, the normalized post-shock density is $\rho_{ps}/\rho_{0} \equiv \rm \mathcal{M}^{2}_{s}$. For roughly virialized clouds, \citet{Burkhart2018} uses this formalism to suggest that the critical density for star formation is approximately the post-shock density for isothermal shocks in supersonic turbulence. In other words, the dynamics of the post-shock regions analyzed in our parameter study may be the most pertinent dynamics for star formation. However, this equation only considers a uniform background density of $\rho_{0}$. In a setup with a uniform background density, the shock experiences a much more coherent density contrast for a shock front of a given Mach number due to uniform jump conditions experienced by the shock. In our simulations, we tend to find density contrasts smaller than the $\mathcal{M}_{s}^{2}$ contrast in the direction of shock propogation due to the warping and corrugation of the shock front. Warping and corrugation also leads to shocks moving into strong turbulence being quickly destroyed, suggesting that the Mach number of the upstream turbulence is an important consideration when applying this equation to star formation analyses.  

The distribution of vorticity in the post-shock region also has implications for the angular momentum of stars that may form there. Several theoretical studies \citep[e.g.][]{Jappsen2004, Chen2018, Misugi2019} have found that initial angular momentum in star-forming cores may be inherited from the surrounding turbulence and yield good agreement with observations \citep{Misugi2019}. Furthermore, \citet{Jappsen2004} found strong correlations between the angular momentum vectors of protostars in close proximity. If transverse vorticity is preferentially selected behind the post-shock region, this may leave some observable signature on the angular momentum of resulting stars. Our vorticity analysis shows that vorticity is present in strong filaments throughout the shock and post-shock regions, suggesting that gas collapsing to form stars in these regions may inherit similar angular momenta. \citet{Pineda2019} find that, even at 1000 AU, the influence of a molecular cloud's turbulence remains present on the angular momentum of a protostar. We propose that the small-scale vorticity generated by a shock propagating through turbulence can help bridge the gap between large-scale turbulence and small-scale protostellar dynamics. Based on the preferential orientation for vorticity in the post-shock region, and if angular momentum is indeed transferred efficiently to smaller scales, this mechanism should lead to an enhancement of star-forming cores or protostellar disks with angular momentum vectors aligned parallel to shock fronts. Future work with magnetic fields and gravity should also shed more light on the potential relationship between post-shock turbulence and the angular momentum of cores.

Furthermore, the generation of converging flows within post-shock regions could form the seeds of star formation or molecular clouds, as theorized by \citet{Burkhart2018} and observed by \citet{Zucker2022_LB} and \citet{Foley2023}. However, additional physics play a large role in moderating the gas motions. Gravity, thermal physics, and magnetic fields all are important to shock evolution. Future work can include these physics to derive a more complete relationship between the dynamics of gas behind shock fronts and star formation.

\section{Conclusions} \label{sec:conclusions}

We have conducted a suite of simulations of a shock moving through turbulence at a variety of Mach numbers. We have found that turbulence develops in the post-shock region of supersonic, compressible turbulence in an anisotropic way, much the same as findings from fluid mechanics simulations of shocks in subsonic turbulence. Namely, we find:

\begin{enumerate}
    \item Turbulent dissipation is strongly enhanced at a shock front and drops off deeper into the post-shock region. Turbulent energy is also significantly raised at a shock front and returns to baseline by the end of the post-shock region. The increase in turbulent energy can almost entirely be attributed to an increase in turbulent fluctuations along the $x$ direction; i.e. in the direction of shock propagation.
    \item A shock front excites a large vorticity magnitude and a strongly negative divergence. Increased vorticity and slightly negative divergences persist throughout the post-shock region. Isolated ``clumps'' of particularly strong vorticity or compression are observed.
    \item Behind the shock, vorticity transverse to the direction of shock propagation is favored. Moving farther into the post-shock region reveals an enhancement of vorticity parallel to the shock propagation, but this is only briefly observed. This means that turbulence just behind a shock can be viewed as roughly a 2D, axisymmetric phenomenon for early times in the shock evolution. By the end of the post-shock region, vorticity will have roughly stabilized. 
    \item Even without gravity, shocks moving through turbulent gas are capable of generating net compressive motions on the scale of star-forming structures, potentially forming the seeds for clumps and clouds in the interstellar medium.
    \item Shock lifetimes are strongly dependent on their movement into decaying or driven turbulence, lasting from 1 Myr to tens of Myr. Decaying turbulence allows shocks to persist for quite some time, while driven turbulence tends to destroy the shock fronts much more quickly.

\end{enumerate}

Shocks moving into supersonic, compressible turbulence follow several of the diagnostics demonstrated in the fluid dynamics community. By studying individual dense regions produced by shock-turbulence interactions, we can gain insight into the physics of the individual components that make up broad statistical measures, such as the column density PDF, and how shocks help facilitate the transition between turbulent and star-forming gas. 

\acknowledgements
We are grateful to the anonymous referee, whose insightful comments improved this manuscript. We wish to thank Eric Koch, Sarah Jeffreson, Vadim Semenov, Theo O'Neill, Ralf Konietzka, and Catherine Zucker for illuminating conversations that shaped the direction of this work. The visualization, exploration, and interpretation of data presented in this work was made possible using the \texttt{glue} visualization software, supported under NSF grant numbers OAC-1739657 and CDS\&E:AAG-1908419. 
\textit{Software:} \texttt{astropy} \citep{AstropyCode}, \texttt{glue} \citep{glueviz2017}, \texttt{numpy} \citep{Numpy2020}, \texttt{PyVista} \citep{PyVista2019}, \texttt{scipy} \citep{Scipy}, \texttt{meshoid} \citep{Grudic2021}

\bibliographystyle{aasjournal}
\bibliography{main}{}


\end{document}